\documentclass[10pt]{article}
\usepackage{amsmath,amsfonts,color}
\usepackage{graphicx}
\usepackage{epsfig,epsf}
\usepackage{hyperref}
\usepackage{bm}

\def\bar {\overline}

\def\bc {B_c}
\def\bs {B_s}

\def\bst {B_s^\ast}

\def\bsbsbar {B_s-\overline{B}_s}

\def\be {\begin{equation}}
\def\ee {\end{equation}}
\def\beq {\begin{equation}}
\def\eeq {\end{equation}}
\def\bea {\begin{eqnarray}}
\def\eea {\end{eqnarray}}

\def\bra {\langle}
\def\ket {\rangle}

\def\beq{\begin{equation}}
\def\eeq{\end{equation}}
\def\barr{\begin{array}}
\def\earr{\end{array}}

\begin{document}

\renewcommand*{\thefootnote}{\fnsymbol{footnote}}

\begin{center}
 {\Large\bf{The role of leptonic cascades in $\bm{B_c}\to \bm{B_s}$ at the LHC}}

\vspace{5mm}

Arghya Choudhury$^{a,b}$\footnote{a.choudhury@sheffield.ac.uk}
Anirban Kundu $^c$ \footnote{anirban.kundu.cu@gmail.com}, and 
Biswarup Mukhopadhyaya $^d$ \footnote{biswarup@hri.res.in}

\vspace{1mm}
$^a$ {\em Consortium for Fundamental Physics, Department of Physics and Astronomy, \\
University of Sheffield, Sheffield S3 7RH, United Kingdom}

$^b$ {\em Consortium for Fundamental Physics, Department of Physics and Astronomy, \\
University of Manchester, Manchester, M13 9PL, United Kingdom}

${}^c$ {\em{Department of Physics, University of Calcutta, \\
92 Acharya Prafulla Chandra Road, Kolkata 700009, India

{$^d$ Regional Centre for Accelerator-based Particle Physics,\\
Harish-Chandra Research Institute, Jhusi, Allahabad 211019, India}
}}

\end{center}
\begin{abstract}
 
We study the feasibility and advantages of investigating the $c\to s$ decay channel of the $B_c$ meson through 
the cascade decay $B_c\to B_s\pi$, $B_s\to D_s\ell\nu$. We take into account possible contaminations coming from 
the vector excitations (like $B_s^*$, $D_s^*$ or $\rho$ in the intermediate stages of the cascade) too, as well as 
the opposite cascade $B_c\to B_s\ell\nu$, $B_s\to D_s\pi$. We show how the $p_T$ distribution of the pion and the 
lepton can possibly help to differentiate between various form factor models, for which we either need more integrated 
luminosity at the LHC, or a high-luminosity $e^+e^-$ B factory producing $B_c$ mesons too. 

\end{abstract}

\date{\today}



\setcounter{footnote}{0}
\renewcommand*{\thefootnote}{\arabic{footnote}}


\rightline{\em\small{Dedicated to the memory of our colleague Debrupa Chakraverty}}

\section{Introduction}

The meson $B_c^{+(-)} \equiv \bar{b} c(b\bar{c})$, the heaviest ground state of an open flavor meson, 
is unique in the sense that both the constituent quarks are heavy and can decay with comparable 
lifetimes, and both $c\to s$ and $b\to c$ decay channels have been observed \cite{pdg,lhcb-1}. 
The excited state $B_c^{*}$ decays promptly to the ground state by a photon emission and 
then follows the usual decay pattern. 
In fact, the $c\to s$ decays (leading to $B_c\to B_s^{(*)}$) 
are more favored than the $b\to c$ decays (leading to $B_c\to J/\psi$ and other charmonia)
because the latter is suppressed by the CKM factor $|V_{cb}|^2$, which more than offsets 
the parent quark mass factor $m_b^5/m_c^5$. About 70\% of the $B_c$ mesons decay via $c\to s$ 
while only 20\% of them decay via $b\to c$ \cite{beneke-prd96,0412158}. 
The annihilation channel $B_c\to \ell\nu$ is not expected to contribute more than 10\%. 
On the other hand, $c\bar{c}$ bound states are 
easier to reconstruct and that is why most of the experimental studies are focussed on them. However, 
there are theoretical studies on the $B_c\to B_s$ decays available in the literature \cite{ckm,galkin,bcbsothers}.

The $\bc \to \bs (\bst)$ decays are important for several reasons. The decays will test, and 
possibly differentiate among, different form factor (FF) models \cite{SWCY}. 
Depending
on the FFs chosen, the branching ratios (BR) can vary by a factor of 5 to 7. For example, the 
BR for $B_c\to B_s\pi$ can vary over a range of $[2.5:16.4]\%$. Apart from differentiating among
various FF models, $B_c\to B_s^{(*)}$ decays
are also the unique testing ground for the effects of a heavy spectator quark. 

In 2013, the LHCb collaboration observed the decay $B_c\to B_s\pi$ \cite{lhcb-1}. The $B_s$, 
in turn, was reconstructed through its hadronic decays $B_s\to D_s\pi$ and $B_s\to J/\psi \phi$, followed by 
 $D_s\to K^+K^-\pi$, $J/\psi\to \mu^+\mu^-$ and $\phi\to K^+K^-$. The vector decay modes like 
 $B_s\to D_s\rho$ or $B_s\to D_s^*\pi$ were treated as pollutants where the soft pion or photon coming 
 from the decay of the vector mesons was not reconstructed. Once the $B_s$ was successfully 
 reconstructed, it was combined with the pion coming at the first stage of the cascade to get the parent 
 $B_c$. Note that $B_c$ can also decay to $B_s^*$, which decays almost with a 100\% BR to $B_s\gamma$. 
 
 The fragmentation function $f_c\equiv {\rm BR}(b\to B_c)$ is, however, not very precisely known at 
 the LHC with $\sqrt{s}=13$ TeV. While the other fragmentation functions are more or less well-measured,
 like $f_u\approx f_d \approx 4 f_s$, the $B_c$ production cross-section $\sigma_{B_c}$ depends crucially on the masses 
 of the $b$ and the $c$ quarks. For example, at $\sqrt{s}=14$ TeV at the LHC, $\sigma_{B_c}$ can vary from 
 28.9 nb to 75.6 nb, for $m_b=4.9$ GeV and $m_c=[1.8:1.3]$ GeV \cite{0309121}. A similar variation is 
 there if one varies $m_b$, which is justifiable from the point that the running $b$-quark mass is between 
 2.5 and 3.0 GeV at the production level because of the higher energy scale involved, whereas for the 
 decay, one should use a higher value of $m_b$ close to 5 GeV. This normalization 
 uncertainty, unfortunately, prevents a precise differentiation among the FF models. To overcome this, 
 at least partially, it is imperative to detect the $B_c\to B_s$ decay, and the subsequent decay of $B_s$, 
 through other final states.

In this paper, we will try to focus on a slightly different cascade decay of $B_c$, leading to a $D_s$, 
a charged pion, a charged lepton, and missing $E_T$ (from the neutrino) in the final state. 
This may or may not be accompanied by one or more photons. The presence of the charged lepton helps 
to make such events relatively clean. The final state can originate through
two distinct chains:
\begin{eqnarray}
 &{\rm Chain~1} : & B_c \to B_s \ell \nu\,,\ \ B_s \to D_s\pi\,,\nonumber\\
 &{\rm Chain~2} : & B_c \to B_s \pi\,,\ \ B_s\to D_s\ell\nu\,.
\end{eqnarray}
Note that we have not shown the electric charges. Without any $\bsbsbar$ oscillation, the lepton 
and the pion would have opposite charges, but the oscillation being quite fast, same sign and opposite 
sign $\pi\ell$ pairs come out with almost equal probability. We have also not shown the vector 
excitations, like $B_c^*$, $B_s^*$, $\rho$, or $D_s^*$. The excitations, in general, come down to the 
pseudoscalar ground states by emitting a pion, if energetically possible, or a photon. Most of the times 
these pions and photons are so soft as to go undetected. An exception is the decay $B_c^+\to B_s\rho^+$
followed by $\rho^+\to \pi^+\pi^0$. The first decay being close to the kinematic threshold, the $\rho$ meson and hence 
the pions tend to
be soft, so one may catch the charged pion and miss the neutral one. 
 It is not easy to remove this channel through 
a mass reconstruction because of the smearing effect in the reconstructed $B_c$ peak; thus, this will be a part of the 
signal, as we discuss later. Events where one sees one or two photons coming out of the $\pi^0$ are vetoed out. 

Naively, we expect many more events in Chain 2 than in Chain 1. The reason is that the three-body decay 
in the first stage
of Chain 1 makes the charged lepton much softer compared to the $\pi^+$ in Chain 2, and so the chance of missing the 
lepton after applying the $p_T$ cuts is quite high. 

We itemize below the motivations for studying the decay cascade of $B_c$ via $B_s$ involving a lepton:  

\begin{itemize}
\item
With a better knowledge of the cascade stage from which the 
pion is coming, one may pin down the FF models more precisely. For example, the relative importance of 
$B_c\to B_s\pi$ and $B_c\to B_s\rho$ depends crucially on the FFs chosen. An example will be given in the next Section. 
Thus, the irreducible contribution of $B_c\to B_s\rho$ on $B_c\to B_s\pi$ is a function of the FFs chosen, and with 
enough data and a clean atmosphere, the momentum distribution of the charged pion may help in differentiating 
among such models, as we will show later. The caveat is that the theoretical uncertainties are yet too high. 


\item
With a lepton in the final state, which is theoretically much cleaner, it should give us a handle 
on the non-negligible $1/m_{b/c}$ effects in the decay, as the $\alpha_s$ corrections are less severe. 

\item
This might turn out to be a good strategy for the future B factories like KEK-B, which will provide 
a much cleaner environment and can also help in the study of possible decay distributions and 
angular correlations of the decay products. If we can somehow reconstruct $B_s^*$, we will have 
more information on several other FFs too. The angular correlation between the pion and the lepton may 
shed light to possible new physics operators present in the decay. 

\item
If we have only one pion in the signal, one may use other decay modes of $D_s$ to reconstruct 
it, with better efficiency, for example $D_s\to 3\pi$ or $D_s\to K\pi\pi$, although the BR of the 
latter mode is much smaller than the other two decay channels. 
\end{itemize}

The paper is arranged as follows. In Section 2, we outline the essential tools for the analysis, while the actual analysis
is performed in Section 3. In the last Section, we summarize and conclude.

\section{The essential principle}

At the LHC, the production cross-section $\sigma_{B_c}$ is of the order of tens of nb, and depends crucially 
on the charm fragmentation function $f_c$. The total cross-section is a sum of the production cross-sections of $B_c$ 
(${}^1S_0$), $B_c^*$ (${}^3S_1$), and other higher excited states. Typical cross-sections for various choices 
of $m_b$ and $m_c$ are shown in Ref.\ \cite{0309121}. We have used the code BCVEGPYv2 \cite{0504017} 
to calculate the production of $B_c$. With $m_b=5.0$ GeV, $m_c=1.27$ GeV, $m_{B_c}=m_b+m_c=6.27$ GeV, and 
the factorization scale $\mu_F^2 = Q^2 = m_{B_c}^2+ p_T^2$ where $p_T$ is the transverse momentum of $B_c$, 
the production cross-section is $\sigma_{B_c}=334$ nb. 
This includes the production of $B_c$, $B_c^*$, and some 
higher resonances, including color-octet $B_c$ states. As they all decay promptly to the ground state $B_c$ by emitting
a photon or a gluon, one must take all of them into account. The gluon distribution function used is CTEQ5L. 
One must note that $\sigma_{B_c}$ depends on the values of $m_b$ and $m_c$, and this normalization uncertainty 
will be inherent in any estimate of the number of events. 

Let us not go into a discussion of any specific theoretical framework; we would refer the reader to Ref.\ \cite{SWCY} 
for a comparative list of different FF models. 
As is well-known, the decay 
$B_c\to B_s$ involves two FFs: 
\be
\bra B_s(p') | V_\mu | B_c (p)\ket =
\frac{m_{B_c}^2 - m_{B_s}^2}{q^2} q_\mu F_0(q^2) + 
\left\{ (p+p')_\mu - \frac{m_{B_c}^2 - m_{B_s}^2}{q^2} q_\mu\right\} F_1(q^2)\,,
\ee
which are equal at zero momentum transfer, $F_0(q^2=0) = F_1(q^2=0)$, with $q=p=p'$. 
However, their forms depend 
on the model chosen to evaluate them. Similarly, $B_c\to B_s^*$ is parametrized by four FFs: 
\bea
\bra B_s^\ast(p',\epsilon) | V_\mu | B_c (p)\ket &=& 
\frac{2V(q^2)}{m_{B_c}+m_{B_s^*}} \epsilon_{\mu\nu\alpha\beta} \epsilon^{\ast\nu} p^\alpha {p'}^\beta\,,\nonumber\\
\bra B_s^\ast(p',\epsilon) | A_\mu | B_c (p)\ket &=& 
i\epsilon^{\ast\nu} 
\left[ 
2m_{B_s^*} A_0(q^2) \frac{q_\mu q_\nu}{q^2} + (m_{B_c} + m_{B_s^*})A_1(q^2) \left(\eta_{\mu\nu} - \frac{q_\mu q_\nu}{q^2} \right) 
\right.\nonumber\\
&& \left. 
-\frac{A_2(q^2)}{m_{B_c}+m_{B_s^*}} q_\nu \left( (p+p')_\mu - \frac{m_{B_c}^2-m_{B_s^*}^2}{q^2} q_\mu \right) \right]\,.  
\eea

To show the model 
dependence, one may mention that $F_0(0)$ for $B_c\to B_s$ can be as low as $0.50$ in relativistic constituent
quark model \cite{galkin,RCQM} or as high as $1.3$ in QCD sum rule based models \cite{QCDS}. The decay width typically goes 
as the square of $F_0$ and can vary by a factor of 6. However, it is not just a question of simple scaling. For example, 
let us take two typical approaches, based on QCD factorization \cite{QCDF} and perturbative QCD \cite{SWCY}. The 
BR of $B_c\to B_s\pi$ is $5.3\times 10^{-2}$ ($8.8\times 10^{-2}$) in the former (latter) approach, while the BR for 
$B_c\to B_s\rho$ is $6.3\times 10^{-2}$ ($3.2\times 10^{-2}$). 
This variation is simply due to the individual FF evaluations in these two models.
While we have not shown the sizable uncertainties in 
these predictions, it is clear that the data, with more precision and a better knowledge of $B_c$ production cross-section
(which essentially is an overall normalization), potentially have the ability to differentiate between the FF models. 

\section{Analysis} 

We simulate $B_c$ production and its subsequent decays by PYTHIAv6.4.28 \cite{pythia} 
coupled with BCVEGPYv2.2 \cite{0504017}. 
Our analysis is in the context of the LHCb detector. 
As we have mentioned, the production cross-section at LHC operating with 
$\sqrt{s}=13$ TeV is 334 nb, including the production of all kinematically allowed higher resonances. The
latter promptly decay 
to the ground state, namely, $B_c (0^-)$. 
We demand a pseudorapidity $2\leq \eta \leq 5$ for $B_c$ and all subsequent particles produced in the cascade. 

For both the decay chains 1 and 2, we have the same final state of one charged lepton, one charged pion, 
one $D_s$, and missing energy, which we will call the $\ell\pi D_s$ final state. 
The $D_s$ is detected primarily through the channel $D_s^+\to K^+K^-\pi^+$ with detection efficiency 
$\epsilon_{D_s}\sim  
85\%$, and this channel also has a large BR, about $5.45\%$. This final state involves one less pion 
than the conventional $B_c\to B_s\pi$, $B_s\to D_s\pi$ cascade, so one can possibly use $D_s \to K\pi\pi$ 
channel too, although this channel is Cabibbo-suppressed. We select the $\pi^+$ and 
$\ell^+$ (where $\ell^+ $ = $e,\mu$) with $2<\eta<5$ and $p_T > 1.5$ GeV 
originating from $B_s$ and $B_c$ for Chain 1 and from $B_c$ and $B_s$ 
for Chain 2 respectively. For $\pi^0$ and $\gamma$ that may come from the vector excitations, 
we keep the same rapidity cut but choose $p_T > 200$ MeV \cite{1412.6352}. The events containing at least 
one photon coming from the decay of $\pi^0$ are vetoed.


Apart from the major detection channel for $D_s$, namely, $D_s^+\to K^+K^-\pi^+$, 
it can also be detected through channels like $K^+\pi^+\pi^-$ and $\pi^+\pi^-\pi^+$. 
The typical detection 
efficiency $\epsilon_{D_s}$ is about 80-90\% \cite{1212.4180} in the former mode and 
that is what we will use in the analysis.  
The mass can be reconstructed from the decay products with 
an accuracy of almost 1\%. Pion detection efficiency is 93\% 
and muon detection efficiency is 97\%, with about 1-3\% 
probability of misidentifying a muon as a pion. 
For momentum measurement, $0.6\%$ uncertainty is a conservative 
estimate; at lower energy, it goes down to $0.4\%$. If $B_s$ 
goes to a muon (which are much easier to detect than the electrons), 
$p_T$ of muon must be greater than 1.48 GeV. For hadronic 
decays, a hadron is required in the calorimeter with $E_T> 3.6$ GeV. 
We also use ${\rm BR}(B_s\to D_s\ell\nu) + {\rm BR}(B_s\to D_s^*\ell\nu) 
= 8.4\%$. 

$B_c^+\to B_s\rho^+$, $\rho^+ \to \pi^+ \pi^0$ can also contribute to the 
same final state if the neutral pion is not detected or missed at the detector. 
$\pi^0$ can be detected from the diphoton invariant mass distribution where 
 $m_{\gamma \gamma}$ is required to be $m_{\pi^0}$ $\pm$ ($\approx$ 25) MeV. 
The photons are demanded to have $p_T >$ 200 MeV.
The $p_T$ cut on $\pi^0$ or photons removes about 62\% of the events coming from intermediate vector excitations. 
The rest add up to the signal events. While a detailed discussion on $\pi^0$ detection is 
available in Ref.\ \cite{pinot}, we take the detection efficiency to be 100\%.

\begin{figure}[h!]
\begin{center} 
\includegraphics[angle =0, width=0.48\textwidth]{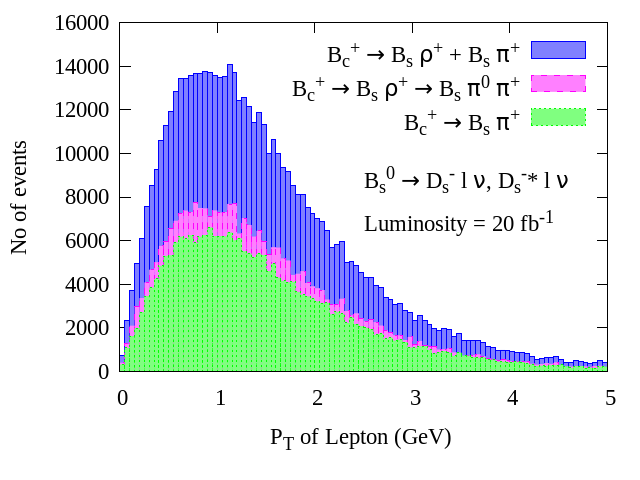}
\includegraphics[angle =0, width=0.48\textwidth]{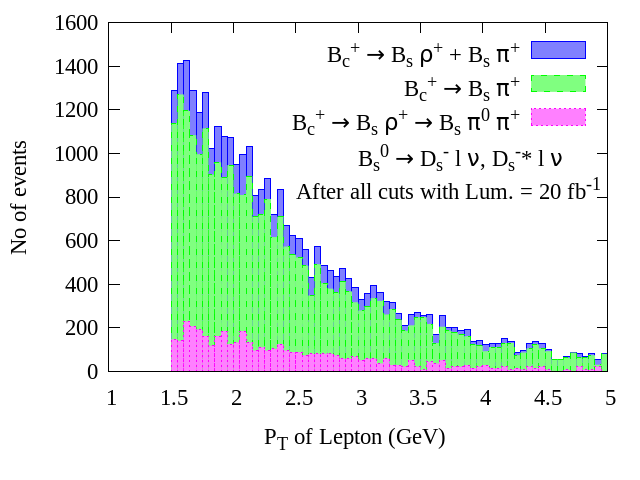}
\includegraphics[angle =0, width=0.48\textwidth]{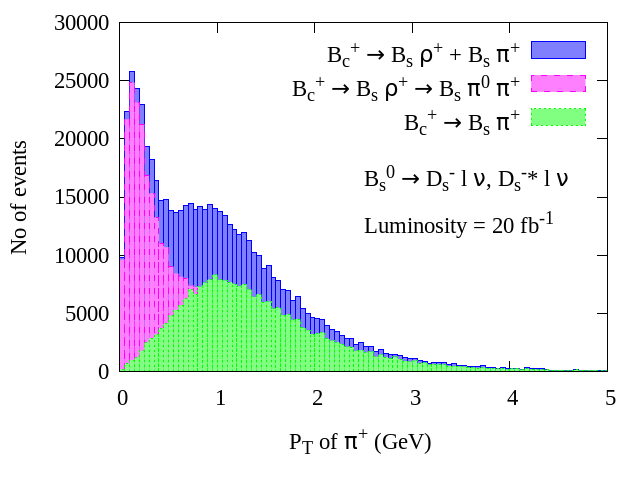}
\includegraphics[angle =0, width=0.48\textwidth]{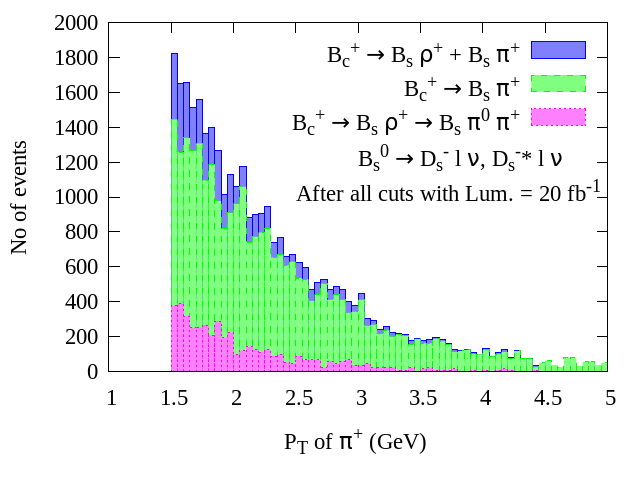}
  \caption{$p_T$ distribution of the lepton (top) and the pion (bottom) before (left) and after (right) 
  the $p_T$ cut and the photon veto for $\pi^0$ are applied. The numbers are only for Chain 2. Chain 1 contributes by about 1\% 
  to these rates because of an undetectably soft lepton in the initial stage of the cascade. 
  }
  \label{fig:ptdist}
    \end{center}
\end{figure}

For the signal, we find, expectedly, that 
the $p_T$ cuts on leptons and charged pions remove most of the events, because at least one of the 
particles is soft enough. 
For $B_c^+\to B_s\pi^+$, about 2.4\% of all events survive, while for $B_c\to B_s\rho$, this is only 0.82\%. 
The latter further reduces to 0.31\% after applying the $p_T$ cuts on $\pi^0$ and/or photons. The distributions are 
shown in Fig.\ \ref{fig:ptdist}.

The number of events passing all the cuts with an integrated luminosity of 20 fb$^{-1}$, $B_c$ production 
cross-section of 334 nb, and pQCD FFs, are as follows:
\bea
B_c\to B_s\ell\nu &=& 514\nonumber\\
B_c\to B_s\pi &=& 29473\nonumber\\
B_c\to B_s\rho &=& 4603\,.
\eea
This implies, as is evident in Fig.\ \ref{fig:ptdist}, that the observed kinematics is essentially that coming from Chain 2.

Our results are shown in Fig.\ \ref{fig:ptdist} for an integrated luminosity of 
$L = 20$ fb$^{-1}$, based on the pQCD FFs. The right-hand plots are with all the cuts applied, including 
the photon veto to remove the identifiable $B_s\rho$ events. The numbers are only for Chain 2; the correction from
Chain 1 is about 1\% after all the cuts are applied, because the lepton coming from a three-body decay of 
$B_c$ is too soft to be detectable most of the times. 

\begin{figure}[h!]
\begin{center} 
\includegraphics[angle =0, width=0.48\textwidth]{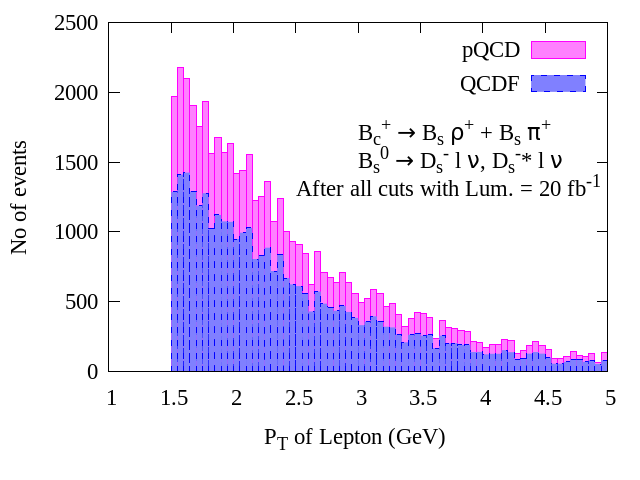}
\includegraphics[angle =0, width=0.48\textwidth]{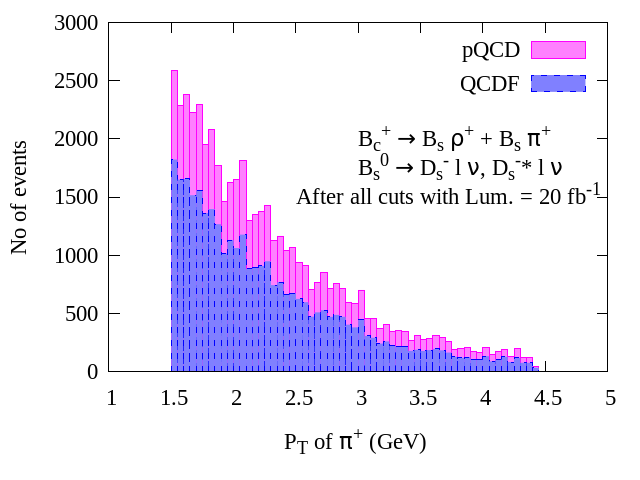}
  \caption{The $p_T$ distribution of pions and leptons in QCDF and pQCD.}
  \label{fig:models}
    \end{center}
\end{figure}

The model dependence of 
the pion or lepton distribution is shown in Fig.\ \ref{fig:models}. We have shown the distributions with two 
FF models, namely, pQCD and QCDF. While the total number of events are quite different, the shapes are almost 
identical, except that for pQCD, the number of comparatively softer pions is slightly more. 
One must mention that the numbers depend on the $B_c$ production cross-section, which in turn depends on 
$f_c$, the $B_c$ fragmentation function. As the latter can vary over a wide range, depending on the values 
of $m_b$ and $m_c$ chosen, one can differentiate among various FF models if the production cross-section, 
which acts as an overall normalization, 
can somehow be fixed.

\section{Summary}

In this paper we have tried to show the possible advantages of looking at the $c\to s$ decay in the $B_c$ meson through 
the cascade $B_c\to B_s\pi$, $B_s\to D_s\ell\nu$. 
This should act as an alternative channel to the cascade where both $B_c$ and $B_s$ decay to pions. 
Once the $B_c$ production cross-section is fixed, the $p_T$ distribution of the charged pion (now that there is no 
confusion from which stage of the cascade it is coming) and the lepton should allow us to guess the relative weightage of 
different decay channels, and hence the corresponding form factors. 

The efficacy of using the channel suggested here, with the appropriate event selection criteria, depends crucially on the 
normalization of the $B_c$ pair production rate. This involves the $B_c$ fragmentation function which in turn is highly sensitive
to $m_b$ and $m_c$. Independent methods of pinning down the uncertainties in this sector will thus be extremely helpful 
in improving our understanding of $B_c$ decay dynamics.

{\em Acknowledgements} -- 
The work of AC is supported by the Lancaster-Manchester-Sheffield  Consortium for Fundamental Physics under 
STFC Grant No.\ ST/L000520/1. 
AK acknowledges Department of Science and Technology, Government of India, and 
Council for Scientific and Industrial Research, Government of India, for support through research projects. 
The work of BM was partially supported by funding
available from the Department of Atomic Energy, Government of India,
for the Regional Centre for Accelerator-based particle Physics (RECAPP). AK also acknowledges the hospitality of 
RECAPP while the project was on.  AC would like to thank  
Xian-You Wang and Xing-Gang Wu for various helpful discussions regarding the BCVEGPY generator.


\begin{thebibliography}{99}

\bibitem{pdg} 
F. Abe {\em et al.}  [CDF Collaboration], Phys.\ Rev.\ D {\bf 58}, 112004
(1998); Phys.\ Rev.\ Lett.\ {\bf 81}, 2432 (1998);\\
T.~Aaltonen {\it et al.} [CDF Collaboration],
  Phys.\ Rev.\ Lett.\  {\bf 100}, 182002 (2008)
  [arXiv:0712.1506 [hep-ex]];\\
  R.~Aaij {\it et al.} [LHCb Collaboration],
  Phys.\ Rev.\ Lett.\  {\bf 109}, 232001 (2012)
  [arXiv:1209.5634 [hep-ex]];\\
  R.~Aaij {\it et al.} [LHCb Collaboration],
  Phys.\ Rev.\ Lett.\  {\bf 108}, 251802 (2012)
  [arXiv:1204.0079 [hep-ex]].

\bibitem{lhcb-1} R.~Aaij {\it et al.} [LHCb Collaboration],
  Phys.\ Rev.\ Lett.\  {\bf 111}, no. 18, 181801 (2013)
  [arXiv:1308.4544 [hep-ex]].

\bibitem{beneke-prd96} M.~Beneke and G.~Buchalla,
  Phys.\ Rev.\ D {\bf 53}, 4991 (1996)
  [hep-ph/9601249].

\bibitem{ckm} D.~Choudhury, A.~Kundu and B.~Mukhopadhyaya,
  Mod.\ Phys.\ Lett.\ A {\bf 16}, 1439 (2001).
  
\bibitem{galkin} D.~Ebert, R.~N.~Faustov and V.~O.~Galkin,
  Eur.\ Phys.\ J.\ C {\bf 32}, 29 (2003)
  [hep-ph/0308149].
  
\bibitem{bcbsothers}
H.~F.~Fu, Y.~Jiang, C.~S.~Kim and G.~L.~Wang,
  JHEP {\bf 1106}, 015 (2011)
  [arXiv:1102.5399 [hep-ph]].\\
S.~Naimuddin, S.~Kar, M.~Priyadarsini, N.~Barik and P.~C.~Dash,
  Phys.\ Rev.\ D {\bf 86}, 094028 (2012).\\
  J.~Sun, Y.~Yang, Q.~Chang and G.~Lu,
  Phys.\ Rev.\ D {\bf 89}, no. 11, 114019 (2014)
  [arXiv:1406.4925 [hep-ph]].\\
  J.~Sun, N.~Wang, Q.~Chang and Y.~Yang,
  Adv.\ High Energy Phys.\  {\bf 2015}, 104378 (2015)
  doi:10.1155/2015/104378
  [arXiv:1504.01286 [hep-ph]].
  
\bibitem{0412158} N.~Brambilla {\it et al.} [Quarkonium Working Group Collaboration],
  hep-ph/0412158.




\bibitem{SWCY} 
J.~Sun, N.~Wang, Q.~Chang and Y.~Yang,
  Adv.\ High Energy Phys.\  {\bf 2015}, 104378 (2015)
  [arXiv:1504.01286 [hep-ph]].


\bibitem{0309121} C.~H.~Chang and X.~G.~Wu,
  Eur.\ Phys.\ J.\ C {\bf 38}, 267 (2004)
  [hep-ph/0309121].

\bibitem{0504017} C.~H.~Chang, J.~X.~Wang and X.~G.~Wu,
  Comput.\ Phys.\ Commun.\  {\bf 174}, 241 (2006)
  [hep-ph/0504017];\\
  C.~H.~Chang, X.~Y.~Wang and X.~G.~Wu,
  Comput.\ Phys.\ Commun.\  {\bf 197}, 335 (2015)
  [arXiv:1507.05176 [hep-ph]].

\bibitem{RCQM} M.~A.~Ivanov, J.~G.~Korner and P.~Santorelli,
  Phys.\ Rev.\ D {\bf 73}, 054024 (2006)
  [hep-ph/0602050].

\bibitem{QCDS} A.~K.~Likhoded and A.~V.~Luchinsky,
  Phys.\ Rev.\ D {\bf 82}, 014012 (2010)
  [arXiv:1004.0087 [hep-ph]].

\bibitem{QCDF} 
J.~f.~Sun, Y.~l.~Yang, W.~j.~Du and H.~l.~Ma,
  Phys.\ Rev.\ D {\bf 77}, 114004 (2008)
  [arXiv:0806.1254 [hep-ph]];\\
  arXiv:1602.07027 [hep-ph].
  
\bibitem{pythia} T.~Sjostrand, S.~Mrenna and P.~Z.~Skands,
  JHEP {\bf 0605}, 026 (2006)
  [hep-ph/0603175].

\bibitem{1212.4180} S.~Blusk [LHCb Collaboration],
  arXiv:1212.4180 [hep-ex].

\bibitem{1412.6352} 
  R.~Aaij {\it et al.} [LHCb Collaboration],
  Int.\ J.\ Mod.\ Phys.\ A {\bf 30}, no. 07, 1530022 (2015)
  [arXiv:1412.6352 [hep-ex]].
  
  \bibitem{pinot} O.~Deschamps, F.~P.~Machefert, M.~H.~Schune, G.~Pakhlova and I.~Belyaev,
  LHCb-2003-091, CERN-LHCb-2003-091;\\
  E.~Govorkova,
  arXiv:1505.02960 [hep-ex].
\end{thebibliography}
\end{document}